\documentclass[%
 reprint,
superscriptaddress,
nofootinbib,
 amsmath,amssymb,
 aps,
prd,
]{revtex4-2}

\usepackage{graphicx}
\usepackage{dcolumn}
\usepackage{bm}
\usepackage{hyperref}
\hypersetup{colorlinks,linkcolor=blue,citecolor=green,urlcolor=red}

\usepackage[caption=false]{subfig}

\begin{document}


\title{Particle production by a relativistic \\
semitransparent mirror of finite size and thickness}

\author{Kuan-Nan Lin}
\email{knlinphy@gmail.com}
\affiliation{Department of Physics and Center for Theoretical Sciences, National Taiwan University, Taipei 10617, Taiwan, R.O.C.}
\affiliation{LeCosPA, National Taiwan University, Taipei 10617, Taiwan, R.O.C.}
\author{Pisin Chen}
\email{pisinchen@phys.ntu.edu.tw}
\affiliation{Department of Physics and Center for Theoretical Sciences, National Taiwan University, Taipei 10617, Taiwan, R.O.C.}
\affiliation{LeCosPA, National Taiwan University, Taipei 10617, Taiwan, R.O.C.}
\affiliation{Kavli Institute for Particle Astrophysics and Cosmology, SLAC National Accelerator Laboratory, Stanford University, Stanford, California 94305, U.S.A.}


\begin{abstract}
Production of massless scalar particles by a relativistic semitransparent mirror of finite transverse size and longitudinal thickness in (1+3)D flat spacetime is studied. The derived particle spectrum formula is applied to two specific trajectories. One is the gravitational collapse trajectory invoked in (1+1)D perfectly reflecting moving mirror literature to mimic Hawking radiation, and the other is the plasma mirror trajectory proposed to be realizable in future experiments. It is found that the finiteness of the transverse size leads to diffraction, while the nontrivial thickness amplifies the production rate. We also estimated the particle yield as $\sim 3000$ in a 20-day data acquisition based on the parameters invoked in the proposed AnaBHEL experiment.
\end{abstract}

\maketitle


\section{\label{Introduction}Introduction}
Quantum radiation from perfectly reflecting point mirrors\footnote{By perfectly reflecting point mirror(s), we mean the point(s) in space where the quantum field is subject to (dynamical) Dirichlet boundary condition(s), i.e., the field vanishes at these point(s).} moving in (1+1)D flat spacetime was first studied by Moore \cite{Moore1970}. Remarkably, analogy of this radiation to Hawking radiation \cite{Hawking1975}, which is a curved spacetime effect, were later made by DeWitt \cite{DeWitt1975}, Fulling and Davies \cite{Fulling1976,Davies1977} by assigning the mirror a certain class of trajectories. This type of radiation emitted by a mirror has been referred to as the Moore effect \cite{Moore1970}, mirror-induced radiation (MIR) \cite{Barton1995}, moving mirror radiation \cite{Hotta2015}, or dynamical Casimir effect (DCE) \cite{Dodonov2020}. Aside from the radiation itself, the corresponding partner particle \cite{Hotta2015} and entanglement entropy \cite{Holzhey1994,Wilczek1993,Bianchi2014,Chen2017,Akal2020} have also been applied to the investigation of black hole information loss paradox.

Moving mirror radiation arises from the interaction between the moving mirror and vacuum fluctuations of the quantum field. Due to the existence of the dynamical boundary condition, the initial vacuum state may be different from the final vacuum state, i.e., $\left|0;\text{in}\right>\neq \left|0;\text{out}\right>$. This leads to different notions of particles for the initial and final observers.

Moving mirror model offers an alternative playground in flat spacetime to effectively investigate various black hole physics by prescribing the mirror a certain class of trajectories. Hence recently, an alternative terminology for the moving mirror models, “accelerated boundary correspondence” (ABC) \cite{Foo2020,Good2020dS}, has been introduced.

\begin{figure}[!htb]
\minipage{0.4\textwidth}
  \includegraphics[width=0.8\linewidth]{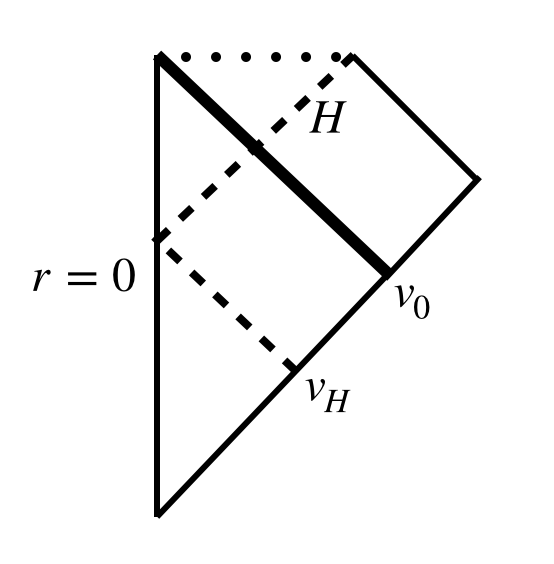}
  \caption{Penrose diagram of gravitational collapse. The flat spacetime patch (lower left) and the Schwarzschild patch (upper right) are glued at $v=v_0$ (thick black line). The event horizon at $r=2M$ is denoted by $H$ (dashed line) and singularity is denoted by the dotted horizontal line.}\label{fig:1}
\endminipage\hfill
\end{figure}

\begin{figure}[!htb]
\minipage{0.4\textwidth}
  \includegraphics[width=0.8\linewidth]{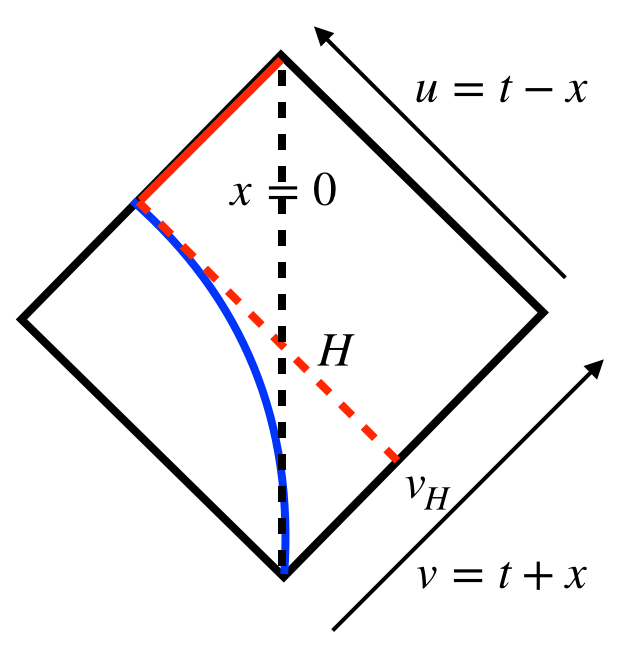}
  \caption{Penrose diagram of a moving mirror in Minkowski coordinates $(t,x)$. The blue curve represents the mirror's trajectory and the red curve represents the horizon.}\label{fig:2a}
\endminipage\hfill
\end{figure}

\begin{figure}[!htb]
\minipage{0.4\textwidth}
  \includegraphics[width=0.8\linewidth]{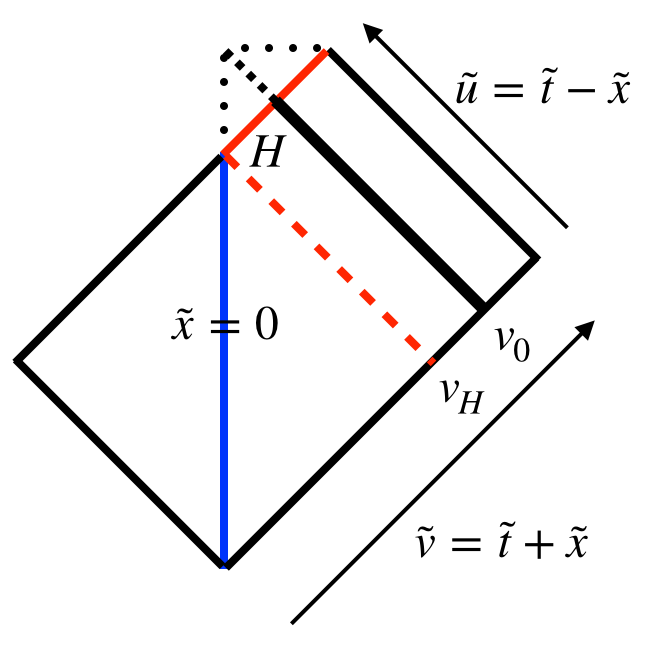}
  \caption{Penrose diagram of a moving mirror in conformal coordinates $(\tilde{t},\tilde{x})$, where $\tilde{u}=p(u),\;\tilde{v}=v$. The region on the mirror's right-hand side resembles that of gravitational collapse (See Fig.~\ref{fig:1}) except that there is nothing inside the future horizon (solid red line). However, the absence of the black hole interior has no effect on the particle production since the production is only relevant to modes intersecting with the mirror's trajectory. On the other hand, the region on the mirror's left-hand side simply resembles that of a flat spacetime.}\label{fig:2b}
\endminipage\hfill
\end{figure}

The correspondence can be easily seen from Figs.~\ref{fig:1}$-$\ref{fig:2b}. Since in the true gravitational collapse scenario, the coordinate origin $(r=0)$ acts effectively as a perfectly reflecting point mirror, conventional moving mirror literature only cares about the right portion of Figs.~\ref{fig:2a} and \ref{fig:2b} and perfectly reflecting boundary condition is imposed on the real mirror itself, which are sufficient to reproduce exactly the same Hawking radiation spectrum. Nonetheless, in laboratory, a real mirror cannot be a perfect reflector. Instead, mirrors are always constructed with finite reflectivities. In this respect, scalar fields may leak through the mirror to the left(right) portion in Figs.~\ref{fig:2a} and \ref{fig:2b} from the mirror's right(left). Although, in this case, the mirror is no longer a perfect reflector, it now acts like an analog wormhole bridging a gravitational collapse spacetime (mirror's right-hand side) and a flat spacetime (mirror's left-hand side).

\begin{figure}[!htb]
\minipage{0.45\textwidth}
  \includegraphics[width=\linewidth]{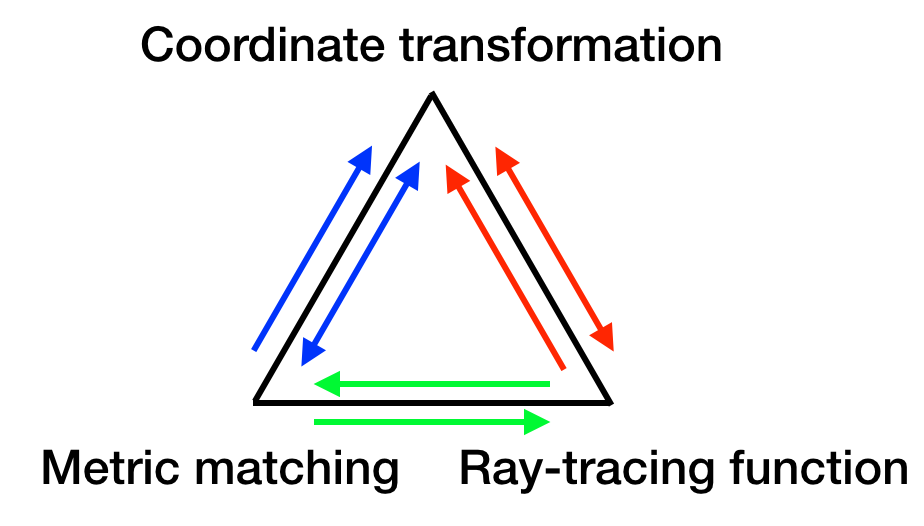}
  \caption{The moving mirror scenario follows the procedure indicated by the inner arrows, while the black hole scenario follows the outer arrows.}\label{fig:triangular relation}
\endminipage\hfill
\end{figure}

Comparisons between the conventional moving mirror and black hole scenarios are summarized in Table~\ref{T1}. In the table, Schwazschild spacetime is used as an example for the black hole scenario, in which $v_0$ is the collapsing null shell's trajectory, $h(u_{\text{Min}},v_0)=h(r)=1-2M/r$ is the metric component, $M$ is the black hole mass, and $f(u_{\text{Min}},v_0)=u_{\text{Sch}}$ is the ray-tracing function which determines the matching of the Minkowski and Schwazschild coordinates. In the moving mirror scenario, $v_0$, $h$, and $u=f(\tilde{u})$ are the corresponding analogs. In addition, note that the plane wave mode that is positive frequency at both null infinities $\mathcal{I}^{-}$ and $\mathcal{I}^{+}$ in the accelerated frame is in fact a positive-frequency plane wave mode at $\mathcal{I}^{-}$ in the lab frame, which together with the positive-frequency plane wave mode at $\mathcal{I}^{+}$ in the lab frame formulates an \textit{in-out} quantum field theory problem in the lab frame. The procedures involved in both scenarios are also illustrated in Fig.~\ref{fig:triangular relation}.

In addition to the analogy of moving mirror model with black hole physics, treating the mirror as a quantum channel that transmits classical and quantum information has been studied in Ref.~\cite{Gianfelici2017} for a perfectly reflecting mirror and in Ref.~\cite{Good2021qc} for a partially reflecting mirror in (1+1)D flat spacetime.

\begin{table*}
\caption{Comparisons between the conventional moving mirror and black hole scenarios in (1+1)D.}
\begin{align}
\begin{array}{c||c||c}
\hline\hline
&\text{Moving mirror scenario (real mirror)} & \text{Black hole scenario (effective mirror)}\\ \hline
\text{Lab frame}&\text{Minkowski: } (u,v)&\text{Schwarzschild: }(u_{\text{Sch}},v)
\\
\text{Lab frame metric}&ds^2=-dudv&ds^2=-h(u_{\text{Min}},v_0)du_{\text{Sch}}dv
\\
\text{Accelerated frame} & \text{Conformal: }(\tilde{u},v) & \text{Minkowski: }(u_{\text{Min}},v)
\\
\text{Accelerated frame metric}&ds^2=-h(\tilde{u},v_0)^{-1}d\tilde{u}dv&ds^2=-du_{\text{Min}}dv
\\
\text{Transformation law}& u=f(\tilde{u},v_0)&u_{\text{Sch}}=f(u_{\text{Min},v_0}) 
\\
\text{Metric vs ray-tracing function}& h(\tilde{u},v_0)^{-1}=df(\tilde{u},v_0)/d\tilde{u} & h(u_{\text{Min}},v_0)^{-1}=df(u_{\text{Min}},v_0)/du_{\text{Min}}
\\
\hline\hline
\end{array} \nonumber
\end{align}
\label{T1}
\end{table*}

Over the years, several trajectories mimicking different black hole scenarios have been studied, including the formation of a Schwarzschild black hole from collapsing null shells \cite{Davies1977}, eternal Schwarzschild black hole \cite{Carlitz1987}, (extremal) Reisnner-Nordstr$\ddot{\text{o}}$m black hole \cite{Good2020ern, Good2020rn}, Kerr-Newman black hole \cite{Foo2020}, dS/AdS spacetime \cite{Good2020dS}, black hole remnants \cite{Good2017}, etc.

Conventional studies of moving mirror model is restricted in (1+1)D flat spacetime since only in this case exact solutions of the mode functions are possible to obtain due to the conformal invariance. However, this technique breaks down in higher dimensions due to the absence of the conformal invariance. Currently, Green function approaches are used to extend the moving mirror model to higher dimensional spacetimes. Perturbations of boundary condition is employed in Refs.~\cite{Ford1982,Neto1996,Rego2013} to extend the model to a non-relativistic, perfectly reflecting, infinite-size plane mirror in (1+3)D; solving the equation of motion (EOM) of the Barton-Calogeracos (BC) action \cite{Barton1995} perturbatively is employed in Ref.~\cite{Lin2020} to extend the model to a relativistic, semitransparent, infinite-size plane mirror in (1+3)D. The latter approach, using an interaction Hamiltonian instead of imposing a strict boundary condition, allows further generalizations of the moving mirror model.

While the mirror-black hole correspondence has been known for decades, there is a lack of experimental support for the generation of the required ultra-relativistic mirror. Recently, an international AnaBHEL (Analog Black hole Evaporation via Lasers) Collaboration \cite{AnaBHEL:2022} has been launched based on the Chen-Mourou \cite{Chen2017prl,Chen2020} proposal aiming at generating the required relativistic mirror via laser-plasma interaction and, ultimately, to study the quantum entanglement between the analog Hawking particle and its partner.

Despite such generated mirror can be relativistic, it inevitably has a low reflectivity \cite{Liu2020}. Let alone the mirror having finite transverse area and longitudinal thickness. In this paper, we aim to study particle creation by the mirror with these issues taken into account.

Our study is based on the following BC-inspired action:
\begin{equation}
\begin{aligned}
S=-\frac{1}{2}\int d^4x\partial^{\mu}\phi(x)\partial_{\mu}\phi(x)
-\frac{\alpha}{2}\int d^4xV(x)\phi^2(x),
\end{aligned}\label{action}
\end{equation}
where $\phi(x)$ is a real scalar field, $V(x)$ is the \textit{potential} that encodes information of the mirror's trajectory, and $\alpha$ is a coupling constant related to the surface density of the charged elements on the mirror with the dimension of length${}^{-1}$ \cite{Barton1995}. Having the action at hand, we may then study various issues such as the particle production, the energy flux, the entanglement entropy, etc. 

This paper is organized as follows. In Sec.~\ref{Bogoliubov coefficients}, we derived the Bogoliubov coefficients for generic $V(x)$. In Sec.~\ref{Applications}, we applied the Bogoliubov coefficients to the gravitational collapse and plasma mirror trajectories to study their corresponding particle spectra and estimate the event yield. Partner particle is discussed in Sec.~\ref{Partner particle}. Conclusion is given in Sec.~\ref{Conclusion}. In Appx.~\ref{A1}, exact solution in (1+1)D is derived. In Appx.~\ref{Transition amplitude}, particle spectrum is alternatively derived via the LSZ formalism. In Appx.~\ref{Diffraction of mode function}, diffraction of field mode is compared to the conventional scalar diffraction theory. In Appx.~\ref{Time-dependent spectrum}, discussion of wave packets is given.

\textbf{Notation:} throughout the paper, we use the mostly plus metric convention, $G=\hbar=c=k_B=1$, $x^\mu=(t,\mathbf{x})$, $\mathbf{x}=(\mathbf{x}_{\perp},z)$, $\mathbf{x}_{\perp}=(x,y)$, and $\mathbb{R}=(-\infty,+\infty)$.

\section{Bogoliubov coefficients} \label{Bogoliubov coefficients}

The equation of motion (EOM) of the BC-inspired action with the observation point $x$ valid in all space is
\begin{equation}
\partial^{\mu}\partial_{\mu}\phi(x)=\alpha V(x)\phi(x),\quad \mu=0,1,2,3,
\end{equation}
where the mirror's information, e.g., dynamics, spatial distribution, etc., is encoded into the potential $V(x)$.

As usual, we are interested in what the observer in the infinite future and on the mirror's right-hand side will observe and we will assume the interaction between the scalar field and the mirror is adiabatically switched on and off in the infinite past and future, respectively, so that the scalar field asymptotes a free field, i.e.,
\begin{equation}
\lim_{t\rightarrow -\infty}\phi(x)=\phi_{\text{in}}(x),\quad\lim_{t\rightarrow +\infty}\phi(x)=\phi_{\text{out}}(x)
\end{equation}
with
\begin{equation}
\begin{aligned}
\phi_{\text{in}}(x)&=\int_{\mathbb{R}}d^2k_{\perp}\int_{\mathbb{R}}dk_3\left[\hat{a}_{\mathbf{k}}^{\text{in}}u^{\text{in}}_{\mathbf{k}}(x)+h.c. \right],
\\
\phi_{\text{out}}(x)&=\int_{\mathbb{R}}d^2k_{\perp}\int_{\mathbb{R}^+}dk_3\left[\hat{a}_{\mathbf{k}}^{\text{out}}v^{\text{out}}_{\mathbf{k}}(x)+h.c. \right],
\end{aligned}
\label{in/out field}
\end{equation}
where
\begin{equation}
u^{\text{in}}_{\mathbf{k}}(x)=v^{\text{out}}_{\mathbf{k}}(x)=\frac{e^{ik\cdot x}}{[2\pi]^{3/2}[2\omega_{k}]^{1/2}}
\end{equation}
are the asymptotic mode functions and $\omega_k=|\mathbf{k}|$, $k\cdot x=-\omega_k t+k_ix^i$, $h.c.$ represents Hermitian conjugate, and we will take the quantum state as the in-vacuum state $\left|0;\text{in}\right>$ defined by $\hat{a}_{\mathbf{k}}^{\text{in}}\left|0;\text{in}\right>=0\;\forall\;\mathbf{k}$.

The solution to the EOM in the case of a semitransparent mirror can be obtained perturbatively as\footnote{Keeping terms to the first order in $\alpha$ is valid for high frequency modes, i.e., semitransparent limit. In this case, the integration domain for the momentum should be determined by some cutoffs with $\alpha$ being the smallest parameter. Thus, we shall replace $\mathbb{R}$ in Eq.~\eqref{in/out field} by $\mathbb{D}_\perp$ for $\mathbf{k}_\perp$ and $\mathbb{D}_3$ for $k_3$ and leave their explicit range until Sec.~\ref{Applications} where concrete examples can be worked with.}
\begin{equation}
\begin{aligned}
\phi&\approx\phi_{\text{in}}(x)-\alpha\int_{\mathbb{R}}d^4x'G_{R}(x,x')V(x')\phi_{\text{in}}(x')\label{Field first order in}
\end{aligned}
\end{equation}
and
\begin{equation}
\begin{aligned}
\phi&\approx\phi_{\text{out}}(x)-\alpha\int_{\mathbb{R}}d^4x'G_{A}(x,x')V(x')\phi_{\text{in}}(x')\label{Field first order out}
\end{aligned}
\end{equation}
to the first order in $\alpha$, i.e., $\phi_{\text{out}}(x)\approx\phi_{\text{in}}(x)+\mathcal{O}(\alpha)$, where
\begin{equation}
G_{R/A}(x,x')=\frac{\delta[t-t'\mp|\mathbf{x}-\mathbf{x}'|]}{4\pi|\mathbf{x}-\mathbf{x}'|}
\end{equation}
are the (1+3)D free field retarded (upper minus sign) and advanced (lower plus sign) Green functions satisfying
\begin{equation}
\partial^{\mu}\partial_{\mu}G_{R/A}(x,x')=-\delta^{(4)}(x-x').
\end{equation}

For the ease of later computations, it would be more convenient to express the retarded and advanced Green functions in momentum space as
\begin{align}
G_{R/A}(x,x')=&2\text{Re}\left[\frac{i}{16\pi^3}\int_{\mathbb{R}^+}d\omega\;e^{\mp i\omega(t-t')}\int_{\mathbb{R}}d^2k_{\perp}\right.\nonumber
\\
&\left.\times e^{i\mathbf{k}_{\perp}\cdot(\mathbf{x}_{\perp}-\mathbf{x}_{\perp}')}\frac{e^{i\sqrt{\omega^2-k_{\perp}^2}|z-z'|}}{\sqrt{\omega^2-k_{\perp}^2}}\right].
\label{R/A green function momentum space}
\end{align}

Taking Fourier transformation of Eqs.~\eqref{Field first order in} and \eqref{Field first order out}:
\begin{equation}
\int_{\mathbb{R}}dt\int_{\mathbb{R}}d^2x_{\perp}\int_{\mathbb{R}^+}dz\;e^{i\omega t-i\mathbf{k}\cdot\mathbf{x}}\hat{\phi}(x),
\end{equation}
using expression \eqref{R/A green function momentum space} in the computations, and imposing conditions: $\omega>0$, $\omega^2>k_{\perp}^2$, and $k_3=\sqrt{\omega^2-k_{\perp}^2}=\sqrt{\omega_{k}^2-k_{\perp}^2}$, we find the Bogoliubov transformation on the mirror's right-hand side as
\begin{equation}
\begin{aligned}
\hat{a}^{\text{out}}_{\mathbf{k}}\approx\hat{a}^{\text{in}}_{\mathbf{k}}+\int_{\mathbb{D}}d^3p\left[\alpha_{\mathbf{kp}}^{(1)}\hat{a}^{\text{in}}_{\mathbf{p}_{++}}+\beta_{\mathbf{kp}}^{(1)}\hat{a}^{\text{in}\dagger}_{\mathbf{p}_{-+}} \right],
\end{aligned}
\end{equation}
where $\mathbf{k}=(\mathbf{k}_{\perp},k_3)$, $\mathbf{p}=(\mathbf{p}_{\perp},p_3)$, $\mathbf{p}_{\pm +}=(\pm\mathbf{p}_{\perp},p_3)$, and the first-order alpha/beta Bogoliubov coefficients for some potential $V(x')$ yet to be specified are 
\begin{widetext}
\begin{align}
\alpha_{\mathbf{kp}}^{(1)}&=-\frac{i\alpha}{16\pi^3\sqrt{\omega_k}\sqrt{\omega_p}}\int_{\mathbb{R}}d^4x'V(x')e^{i(\omega_k-\omega_p)t'-i(k_3-p_3)z'}e^{-i(\mathbf{k}_{\perp}-\mathbf{p}_{\perp})\cdot\mathbf{x}_{\perp}'},\label{alpha}
\\
\beta_{\mathbf{kp}}^{(1)}&=-\frac{i\alpha}{16\pi^3\sqrt{\omega_k}\sqrt{\omega_p}}\int_{\mathbb{R}}d^4x'V(x')e^{i(\omega_k+\omega_p)t'-i(k_3+p_3)z'}e^{-i(\mathbf{k}_{\perp}+\mathbf{p}_{\perp})\cdot\mathbf{x}_{\perp}'}.\label{beta}
\end{align}
\end{widetext}

The number of particles emitted to the mirror's right-hand side per $\mathbf{k}$ mode in the infinite future is thus
\begin{equation}
\begin{aligned}
\frac{dN}{d^3k}&=\left<0,\text{in}\right|\alpha_{\mathbf{k}}^{\text{out}\dagger}\alpha_{\mathbf{k}}^{\text{out}}\left|0,\text{in}\right>
\approx\int_{\mathbb{D}}d^3p\;|\beta^{(1)}_{\mathbf{kp}}|^2,
\end{aligned}\label{Number of particles}
\end{equation}
where $p_3<0$ is the contribution from particles created by reflected in-mode and $p_3>0$ is the contribution from particles created by transmitted in-mode.

\subsection{Infinite-size plane mirror}\label{An infinite-size plane mirror}

The original BC action models an infinite-size plane mirror of zero thickness with the potential given by \cite{Barton1995}
\begin{equation}
V(x)=\gamma^{-1}(t)\delta[z-Z(t)],\label{Infinite plane mirror}
\end{equation}
where we have let the mirror move in the 3-direction.

Inserting potential \eqref{Infinite plane mirror} into the beta coefficient \eqref{beta}, we obtain
\begin{align}
\beta_{\mathbf{kp}}^{(1)}&=\frac{-i\alpha}{16\pi^3\sqrt{\omega_k}\sqrt{\omega_p}}\;(2\pi)^2\delta(\mathbf{k}_{\perp}+\mathbf{p}_{\perp})\nonumber\\
&\quad\times\int_{\mathbb{R}}dt\gamma^{-1}(t)e^{i(\omega_k+\omega_p)t-i(k_3+p_3)Z(t)},\label{beta infinite plane}
\end{align}
which reproduces the result in our previous work \cite{Lin2020}.

\subsection{Finite-size plane mirror}\label{A finite-size plane mirror}

Let us next consider the case of a plane mirror moving in the 3-direction with a finite size: $A=L\times L$, where $L$ is the side width, centered at $(x,y)=(0,0)$ in the transverse dimensions. The potential that models this kind of mirror is
\begin{equation}
V(x)=\gamma^{-1}(t)\delta[z-Z(t)]H(x,y),\label{Finite plane mirror}
\end{equation}
where
\begin{equation}
\begin{aligned}
H(x,y)=&\left[\Theta(x+L/2)-\Theta(x-L/2)\right]
\\
&\times\left[\Theta(y+L/2)-\Theta(y-L/2)\right]
\end{aligned}
\end{equation}
with $\Theta(x)$ being the Heaviside step function, models the mirror's transverse distribution.

Inserting potential \eqref{Finite plane mirror} into the beta coefficient \eqref{beta}, we obtain
\begin{align}
\beta_{\mathbf{kp}}^{(1)}&=\frac{-i\alpha A}{16\pi^3\sqrt{\omega_k}\sqrt{\omega_p}}\;\text{sinc}\frac{(k_1+p_1)L}{2}\;\text{sinc}\frac{(k_2+p_2)L}{2}\nonumber
\\
&\quad\times\int_{\mathbb{R}}dt\gamma^{-1}(t)e^{i(\omega_k+\omega_p)t-i(k_3+p_3)Z(t)},\label{beta finite plane}
\end{align}
where $A=L^2$ is the mirror's area and $\text{sinc}(x)=\sin(x)/x$. The prefactors being sinc functions instead of delta functions due to the finiteness of the mirror's transverse dimensions, which results in the diffraction of field modes.

In the case of large side width, i.e., $L\rightarrow\infty$, the following identity:
\begin{equation}
\lim_{L\rightarrow\infty}\text{sinc}\frac{(k_i+p_i)L}{2}=\frac{2\pi}{L}\delta(k_i+p_i)
\end{equation}
successfully transforms Eq.~\eqref{beta finite plane} into Eq.~\eqref{beta infinite plane}.

\subsection{Finite-size SRLD mirror}\label{A finite-size SRLD mirror}

A mirror of finite transverse size with square-root-Lorentzian distribution (SRLD) in the longitudinal direction may be more relevant to real experiments. This mirror can be modeled by
\begin{equation}
V(x)=\gamma^{-1}(t)f[z-Z(t)]H(x,y),\label{Finite plane mirror}
\end{equation}
where
\begin{equation}
\begin{aligned}
f[z-Z(t)]=\frac{1}{\sqrt{(z-Z(t))^2+W^2}},
\end{aligned}
\end{equation}
where $W$ is the half width at half maximum of the square-root-Lorentzian distribution.

In this case, we have
\begin{align}
&\beta_{\mathbf{kp}}^{(1)}=\frac{-i\alpha A}{16\pi^3\sqrt{\omega_k}\sqrt{\omega_p}}\;\text{sinc}\frac{(k_1+p_1)L}{2}\;\text{sinc}\frac{(k_2+p_2)L}{2}\nonumber
\\
&\times 2K_{0}(W|k_{3}+p_{3}|)\int_{\mathbb{R}}dt\gamma^{-1}(t)e^{i(\omega_k+\omega_p)t-i(k_3+p_3)Z(t)},
\label{beta finite SRLD}
\end{align}
where $K_{0}$ is the modified Bessel function of the second kind of order 0. The appearance of $K_{0}$ indicates the possibility of amplifying the particle production rate through the mirror's nontrivial longitudinal distribution.

\section{Applications}\label{Applications}
\subsection{Trajectory for analog gravitational collapse}

We now study the particle production by a semitransparent plane mirror of finite transverse size. In particular, we consider the trajectory \cite{Nicolaevici2009}:
\begin{equation}
\begin{aligned}
Z(t)
=
\begin{cases}
0\;,\quad t\leq 0
\\
-t+\frac{1-W[e^{1-2\kappa t}]}{\kappa}\;,\quad 0\leq t<\infty,
\end{cases}
\end{aligned}
\label{trajectory 1}
\end{equation}
where $W(x)$ is the product logarithm and $\kappa$ is a positive parameter. In the moving mirror picture, this trajectory models a mirror that is initially at rest but accelerates leftward to asymptotically null at late time (the null line asymptoted by the trajectory is the analog of past event horizon in the black hole spacetime). In the black hole picture, this trajectory models the collapse of a null shell that forms a black hole at late time with $\kappa$ mimicking the black hole's surface gravity.

By inserting trajectory \eqref{trajectory 1} into the beta coefficient \eqref{beta finite plane} and following the calculation procedures outlined in Ref.~\cite{Lin2020}, one obtains, for a finite-size plane mirror,
\begin{widetext}
\begin{align}
\beta_{\textbf{k}\textbf{p}}^{(1),ref}&\approx\frac{A}{4\pi^2}\;\text{sinc}\left[\frac{(k_1+p_1)L}{2}\right]\text{sinc}\left[\frac{(k_2+p_2)L}{2}\right]\left[\frac{-\alpha}{4\pi \sqrt{\omega_{k}}\sqrt{\omega_{p}}} \right]
\label{beta trajectory 1 exact}
\\
&\quad\times\left\{\frac{1}{\omega_{k}+\omega_{p}}+\frac{1}{\kappa}\;\text{exp}\left(\frac{i\pi}{4}+\frac{i\omega_{-}^r}{2\kappa}-\frac{\pi \omega_{+}^r}{4\kappa}\right)\left(\frac{\omega_{-}^r}{2\kappa}\right)^{-\frac{1}{2}+i\omega_{+}^r/2\kappa}
\biggl[\Gamma\left(\frac{1}{2}-\frac{i\omega_{+}^r}{2\kappa} \right)-\Gamma\left(\frac{1}{2}-\frac{i\omega_{+}^r}{2\kappa},\frac{i\omega_{-}^r}{2\kappa}\right) \biggr]\right\},
\nonumber
\end{align}
\end{widetext}
where $p_3>0$\footnote{We have made a change of notation: $p_{3}\to -p_{3}$, where $p_{3}<0$ on the lhs and $p_{3}>0$ on the rhs.}, $\omega_{-}^r=  \omega_{k}-k_3+\omega_{p}+p_3$, and $\omega_{+}^r= \omega_{k}+k_3+\omega_{p}-p_3$, for the particles emitted to the mirror's right-hand side due to the reflected in-mode\footnote{Particle creation due to the transmitted in-mode or particle creation on the left-hand side of the mirror can also be computed. However, we only consider particle creation due to the reflected in-mode on the mirror's right-hand side in this paper since it plays the role of analog Hawking radiation.}. Since Eq.~\eqref{beta trajectory 1 exact} is a first-order result, it is only valid for $\alpha\ll\kappa$, $\alpha\ll\omega_{k}$, $\alpha\ll\omega_{p}$. In addition, since the mirror is moving in the $3$-direction, it is intuitively to restate $\alpha\ll\omega_{k}$ and $\alpha\ll\omega_{p}$ as $\alpha\ll k_3$ and $\alpha\ll p_3$, which naturally leads to $\alpha\ll\omega_{k}$, $\alpha\ll\omega_{p}$. In terms of spherical coordinates, the conditions: $\alpha\ll k_3$ and $\alpha\ll p_3$ are equivalent to $\theta_{k}\ll\pi/2$ and $\theta_{p}\ll\pi/2$, where $k_3=\omega_k\cos\theta_k$ and $p_3=\omega_p\cos\theta_p$.

(Analog) Hawking radiation is a late time phenomenon since it originates from vacuum fluctuations near the horizon, which emerges at late time. In addition, since the mirror is asymptotically null at late time, only incident modes with almost-vanishing transverse momenta on the mirror's right-hand side can catch up the receding mirror to get reflected to experience significant Doppler redshift in frequency. These suggest that the analog Hawking radiation is related to $\mathbf{k}$ and $\mathbf{p}$ in the following regime: $\omega_{p}\sim p_3$ and $\omega_{p}\gg\omega_{k}$. On one hand, in the semiclassical black hole picture, one requires the mass of the black hole $M$ to be larger than the energy of the particle's. This leads to $\omega_{k}\gg\kappa$ since $\kappa\sim1/M$. On the other hand, quantum fluctuations with $\omega_k\lesssim\kappa$ can also give rise to the Hawking radiation according to the generalized uncertainty principle (GUP) \cite{Adler2001}. Motivated by the above mentioned, we study the beta coefficient in the limits of $\omega_{p}\sim p_3,\;\omega_{p}\gg\omega_{k},\;\omega_{p}\gg\kappa$ and obtain the modulus squared of Eq.~\eqref{beta trajectory 1 exact} as
\begin{equation}
\begin{aligned}
&|\beta_{\textbf{k}\textbf{p}}^{(1),ref}|^2
\\
&\approx\frac{A^2}{16\pi^4}\;\text{sinc}^2\left[\frac{(k_1+p_1)L}{2}\right]\text{sinc}^2\left[\frac{(k_2+p_2)L}{2}\right]
\\
&\quad\times
\frac{\alpha^2}{8\pi\kappa\omega_{k} ({p_3})^2}\left[\frac{1}{e^{\omega_{k}/T_{\text{eff}}(\theta_k)}+1}\right],
\end{aligned}\label{beta squared trajectory 1}
\end{equation}
where we have identified the effective temperature by $T_{\text{eff}}(\theta_k)=\kappa/[(1+\cos\theta_k)\pi]$. In the direction normal to the mirror's surface, i.e., $\theta_k=0$, the emitted particles mimic the Hawking radiation with the conventional Hawking temperature $T_{\text{eff}}(0)=T_H=\kappa/2\pi$.

The corresponding particle spectrum, i.e., number of particles per frequency per unit solid angle, is
\begin{equation}
\begin{aligned}
&\frac{dN_{ref}}{d\omega_{k} d\Omega}\approx\omega_{k}^2\int_{\mathbb{D}_3^{+}} d{p}_3\int_{\mathbb{D}_\perp} d^2p_{\perp}\;|\beta_{\textbf{k}\textbf{p}}^{(1),ref}|^2
\\
&\approx\frac{\alpha^2}{8\pi\kappa}\left[\frac{\omega_{k}}{e^{\omega_{k}/T_{\text{eff}}(\theta_k)}+1}\right]\int_{\mathbb{D}_3^{+}} dp_{3}\frac{\mathcal{F}_{L}(\mathbf{k}_{\perp},P_{\perp})}{(p_3)^2},
\end{aligned}\label{Particle spectrum trajectory 1}
\end{equation}
where $\mathbb{D}_{\perp}=(-P_i,P_i)$, $i=1,2$, $\mathbb{D}_{3}^{+}=(P_3,\infty)$. $P_i$ is introduced since $\omega_{p}\sim p_3$; $P_3$ is the infrared cutoff for $p_3$ required by $p_3\gg \alpha$, $p_3\gg \omega_{k}$, and $p_3\gg \kappa$, whereas the finite-size effect is encoded in the form factor:
\begin{widetext}
\begin{align}
\mathcal{F}_{L}(\mathbf{k}_{\perp},P_{\perp})&=\frac{A^2}{16\pi^4}\int_{-P_{\perp}}^{P_{\perp}}dp_{1}\text{sinc}^2\left[\frac{(k_1+p_1)L}{2}\right]f_L(k_2,p_1,P_{\perp}),\label{FL}
\\
f_L(k_2,p_1,P_{\perp})&=\int_{-\sqrt{P_{\perp}^2-p_1^2}}^{\sqrt{P_{\perp}^2-p_1^2}}dp_2\;\text{sinc}^2\left[\frac{(k_2+p_2)L}{2}\right]\nonumber
\\
&=\frac{2}{L}\biggl[ \text{Si}[(k_2+\sqrt{P_{\perp}^2-p_1^2})L]-\text{Si}[(k_2-\sqrt{P_{\perp}^2-p_1^2})L] \biggr]\label{fL}
\\
&\quad-\frac{4\sqrt{P_{\perp}^2-p_1^2}}{L^2(P_{\perp}^2-p_1^2-k_2^2)}\biggl[1-\cos(k_2 L)\cos(\sqrt{P_{\perp}^2-p_1^2} L)-\frac{k_2}{\sqrt{P_{\perp}^2-p_1^2}}\sin(k_2 L)\sin(\sqrt{P_{\perp}^2-p_1^2} L) \biggr],\nonumber
\end{align}
\end{widetext}
where $P_{\perp}=p_{3}\tan\theta_{*}$, $\theta_{*}$ is the maximum polar angle $\theta_{p}$ can be due to the condition $\omega_p\sim p_3$, i.e., $\theta_{p}\leq \theta_{*}\ll 1$, and $\text{Si}(x)$ is the sine integral defined by
\begin{equation}
\text{Si}(x)=\int_{0}^{x}dy\;\text{sinc}(y)\end{equation}
with the special values: $\lim_{x\rightarrow\pm\infty}\text{Si}(x)=\pm\pi/2$.

\begin{figure}[!htb]
\minipage{0.4\textwidth}
  \includegraphics[width=\linewidth]{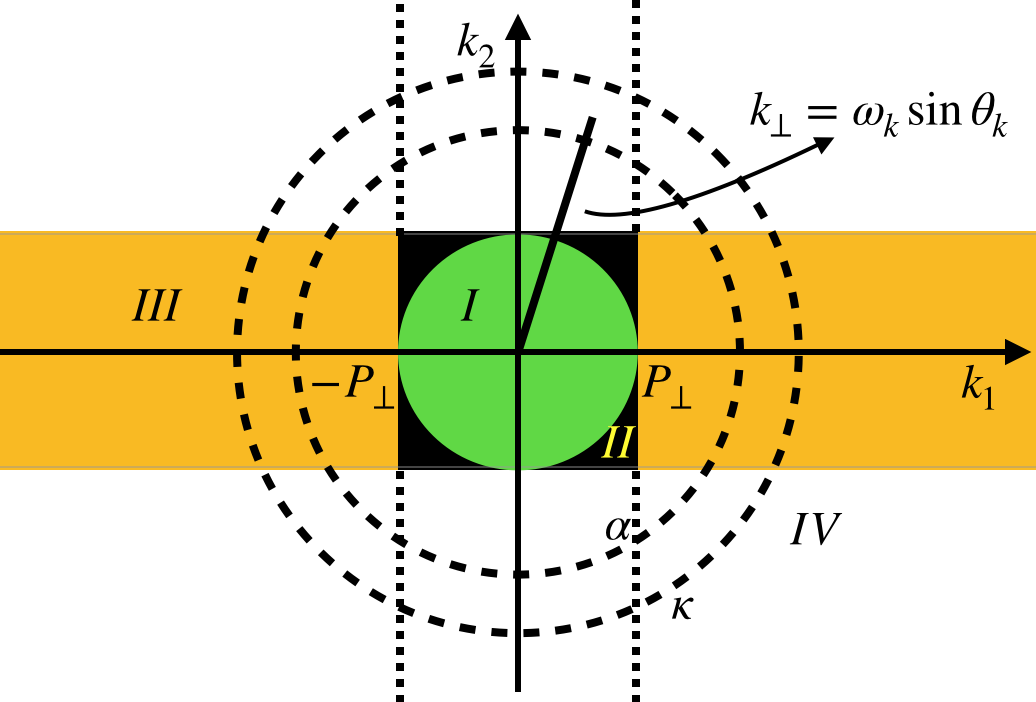}
  \caption{Domain of transverse momenta for analog Hawking particles. Green (Region I): $k_1^2+k_2^2< P_{\perp}^2$; Black (Region II): $|k_1|<P_{\perp},|k_2|<P_{\perp}$, and $\theta_{k}>\arcsin(P_{\perp}/\omega_{k})$; Yellow (Region III): $|k_1|>P_{\perp},|k_2|<P_{\perp}$; White (Region IV): $|k_2|>P_{\perp}$; Inner dashed circle: radius $\alpha$; Outer dashed circle: radius $\kappa$; Solid black line: magnitude of transverse momentum.}\label{fig:domain}
\endminipage\hfill
\end{figure}

\begin{figure*}
\minipage{0.8\textwidth}
\centering
\subfloat{\label{freq1}\includegraphics[width=0.46\textwidth]{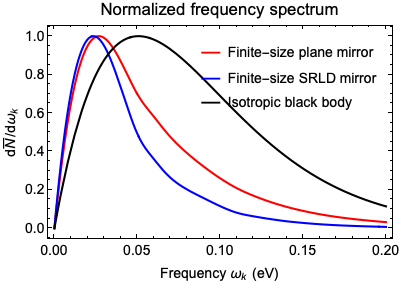}}\qquad
\subfloat{\label{angular1}\includegraphics[width=0.45\textwidth]{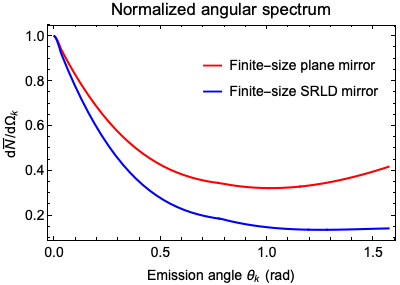}}
\caption{Normalized frequency and angular spectra with temperature $=0.031\text{ eV}$ $(369K)$ in the far infrared regime. In the low frequency regime, $\omega_{k}\rightarrow 0$, the frequency spectra are linear in $\omega_{k}$. Red: $\alpha=0.096\text{ eV},\kappa=0.2\text{ eV}, L=254\text{ eV}^{-1}, \theta_{*}=0.01\text{ rad}, P_{3}=0.2\text{ eV}$. Blue: $\alpha=0.096\text{ eV},\kappa=0.2\text{ eV}, L=254\text{ eV}^{-1}, \theta_{*}=0.01\text{ rad}, P_{3}=0.2\text{ eV}, W=0.0074\text{ eV}^{-1}$. Black: a conventional ideal, isotropic, black body with temperature $\kappa/(2\pi)$, where $\kappa=0.2\text{ eV}$. By saying frequency, we actually mean energy, i.e., $\omega_{k}=2\pi\nu$, where $\nu$ is the actual frequency.}
\label{freqangular}
\endminipage
\end{figure*}

\begin{figure*}
\minipage{0.8\textwidth}
\centering
\subfloat{\label{freq2}\includegraphics[width=0.46\textwidth]{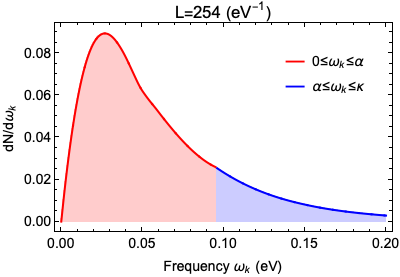}}\qquad
\subfloat{\label{freq3}\includegraphics[width=0.45\textwidth]{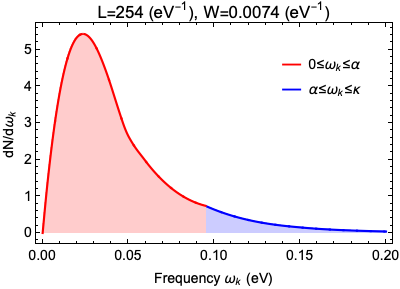}}
\caption{Left: frequency spectrum for finite transverse area, zero thickness, plane mirror. Right: frequency spectrum for finite transverse area, SRLD mirror. The values of parameters used are the same as the previous figure.}
\label{freq}
\endminipage
\end{figure*}

The exact integration in Eq.~\eqref{FL} is complicated. Even the results obtained under certain approximations for simplifications are also lengthy. Thus, we shall omit the analytic results but just simply point out the relevant physics behind the following figures obtained from numerical integrations. However, computation for $L\rightarrow\infty$\footnote{Strictly speaking, $|k_1|L\gg1$, $|k_2|L\gg 1$ for $k_1,k_2\neq 0$.} is straightforward and it helps to develop a sense about the behavior of the form factor. By utilizing the special values of the sine integral for large arguments, we find
\begin{equation}
\begin{aligned}
\mathcal{F}_{\infty}(\mathbf{k}_{\perp})&=
\begin{cases}
\frac{A}{4\pi^2},\quad& \mathbf{k}_{\perp}\in\text{ Region I}
\\
0,\quad&\text{otherwise},
\end{cases}
\end{aligned}
\label{Form factor infinite}
\end{equation}
where Region I refers to the green region in Fig.~\ref{fig:domain} and otherwise refers to regions other than Region I. If $L=$ finite, then $\mathbf{k}_{\perp}$ with values beyond Region I (via diffraction) while $\mathbf{p}_{\perp}\in$ Region I can lead to non-vanishing contribution to the form factor.

The form factor \eqref{Form factor infinite} indicates that the emission of analog Hawking particles is independent of the azimuthal angle $\phi_k$, where $\phi_k=\arctan(k_2/k_1)$, which is expected from translational invariance. In addition, the condition: $\mathbf{k}_{\perp}\in\text{ Region I}$ sets the constraint: $\omega_{k}\sin\theta_{k}<P_{\perp}$, which implies, for a given $P_{\perp}$, the larger the frequency $\omega_{k}$ is, the narrower the emission angle $\theta_{k}$ becomes.

If the mirror has a square-root-Lorentzian distribution (SRLD) in the longitudinal direction, then there will be an additional nontrivial form factor:
\begin{equation}
    \mathcal{F}_{W}(k_{3},p_{3})=\left[2K_{0}(W|k_{3}-p_{3}|)\right]^2,
    \label{Form factor SRLD}
\end{equation}
where $p_{3}>0$, inside the integrand of Eq.~\eqref{Particle spectrum trajectory 1}. That is, the spectrum for this case is
\begin{equation}
\begin{aligned}
&\frac{dN_{ref}}{d\omega_{k} d\Omega}\approx\omega_{k}^2\int_{\mathbb{D}_3^{+}} d{p}_3\int_{\mathbb{D}_\perp} d^2p_{\perp}\;|\beta_{\textbf{k}\textbf{p}}^{(1),ref}|^2
\\
&\approx\frac{\alpha^2}{8\pi\kappa}\left[\frac{\omega_{k}}{e^{\omega_{k}/T_{\text{eff}}(\theta_k)}+1}\right]\int_{\mathbb{D}_3^{+}} dp_{3}\frac{\mathcal{F}_{L}(\mathbf{k}_{\perp},P_{\perp}) \mathcal{F}_{W}(k_{3},p_{3})}{(p_3)^2}.
\end{aligned}\label{Particle spectrum trajectory 1 SRLD}
\end{equation}

In Eq.~\eqref{Particle spectrum trajectory 1 SRLD}, since $p_{3}\gg \omega_{k}$, $k_{3}$ is negligible in Eq.~\eqref{Form factor SRLD}. In addition, let us denote $\bar{p}_{3}$ as the value of $p_{3}$ such that $K_{0}(W\bar{p}_{3})=0$. Then $K_{0}(Wp_{3})>1$ for $p_{3}<\bar{p}_{3}$, and $K_{0}(Wp_{3})<1$ for $p_{3}>\bar{p}_{3}$. This indicates that particle production can be enhanced from the domain $p_{3}<\bar{p}_{3}$. Although the modified Bessel function with $p_{3}>\bar{p}_{3}$ suppresses the production probability, it has negligible effect since, aside from $K_{0}$, the integrand already decays when $p_{3}$ is large whether the mirror has a thickness or not. Thus, overall speaking, a SRLD mirror can give a higher yield of analog Hawking particles compared to a mirror without thickness.

Frequency and angular spectra according to Eqs.~\eqref{Particle spectrum trajectory 1}, \eqref{FL}, \eqref{fL}, \eqref{Form factor SRLD}, and \eqref{Particle spectrum trajectory 1 SRLD} are plotted in Figs.~\ref{freqangular}$-$\ref{freq}. Despite Eq.~\eqref{Particle spectrum trajectory 1} and \eqref{Particle spectrum trajectory 1 SRLD} being derived under the approximation $\alpha\ll\omega_k$, we plot the spectra for $\omega_k\ll\alpha$ using Eq.~\eqref{Particle spectrum trajectory 1} and \eqref{Particle spectrum trajectory 1 SRLD} since it should still remain valid in this regime (See Appx.~\ref{A1}). In addition, parameter values are chosen so as to generate a feasible mirror accelerated by wakefields in the plasma background based on particle-in-cell (PIC) simulation \cite{Liu2020, Liu202105}.

The right panel of Fig.~\ref{freqangular} shows that the emission of particles rises up again for large angles. This can be explained by the conventional Einstein's mirror \cite{Einstein1905,AA2008}. For a non-relativistic, perfectly reflecting mirror without thickness, the incident angle of a photon is equal to its reflected angle. However, if the mirror is receding relativistically, then incident photons with even just a little transverse momenta will be reflected to large angles, which is the case for our analog Hawking particles. 

In general, if the mirror is not a perfect reflector, then its reflectivity is nontrivial in the sense that the reflectivity will be a function of both spacetime coordinates and momentum. The nontrivial reflectivity contributes nontrivially to the beta coefficient and thus alters the particle spectrum. Another factor that alters the particle spectrum is the form factors due to the mirror's geometry, e.g., finite transverse size and thickness. Comparisons of the frequency spectrum for finite-size plane/SRLD semitransparent mirrors to a typical isotropic black body are shown in Fig.~\ref{freqangular}. The curves are normalized with respect to their respective peak values.

Estimation of particle number is important for experimental designs. With the parameter values given in Fig.~\ref{freqangular}, we obtained the number of analog Hawking particles (per laser shot) as $6.5\times 10^{-3}$ for a finite-size plane mirror (left panel of Fig.~\ref{freq}) and $0.29$ for a finite-size SRLD mirror (right panel of Fig.~\ref{freq}). Specifically, $6.5\times 10^{-3}=(5.4+1.1)\times 10^{-3}$ and $0.29=0.27+0.02$, where $5.4\times 10^{-3}$ and $0.27$ refer to the areas shaded in red in Fig.~\ref{freq}, etc. Compared to the mirror without thickness, the SRLD mirror indeed has a higher event yield.

\subsection{Chen-Mourou trajectory}

We now consider the trajectory proposed to be realizable in a future flying plasma mirror experiment \cite{Chen2020}. The trajectory is given by
\begin{equation}
\begin{aligned}
t(Z)
=
\begin{cases}
-\frac{Z}{v},\quad v\rightarrow 1, \;0\leq Z< \infty
\\
-Z+\frac{3\pi}{2\omega_{p0}(1+b)}\left[\frac{1+b}{1+be^{Z/D}}-1 \right], -\infty< Z\leq 0,
\end{cases}
\end{aligned}\label{trajectory 2}
\end{equation}
where $\{\omega_{p0},b,D\}$ are positive plasma mirror parameters and time $t$ is written as a function of the trajectory $Z$. This trajectory is initially moving to the left ultra-relativistically and it approximates the Davies-Fulling trajectory either (i) in a near-uniform plasma background $b\ll 1$, or (ii) at \textit{late time} $t\gg D\ln \left[3\pi/(2\omega_{p0}D)\right]+3\pi/[2(1+b)\omega_{p0}]$ for any $b$. For suitably chosen parameter values and experimental design, this \textit{late time} period may occupy most of the acceleration phase.

In the first case with $b\ll1$, $b\ll\omega_{p0}D$, $\omega_{k}+\omega_{p}\ll\omega_{p0}b^{-2}$ \cite{Lin2020}, or in the second case with the entire acceleration phase dominated by the \textit{late time} period and $b=1$, one obtains, in either cases and for a plane mirror,
\begin{widetext}
\begin{equation}
\begin{aligned}
\beta_{\textbf{k}\textbf{p}}^{(1),ref}&\approx\frac{A}{4\pi^2}\;\text{sinc}\left[\frac{(k_1+p_1)L}{2}\right]\text{sinc}\left[\frac{(k_2+p_2)L}{2}\right]\left[\frac{-\alpha}{4\pi \sqrt{\omega_{k}}\sqrt{\omega_{p}}} \right]\frac{\sqrt{2D}}{\sqrt{\omega_{k}+\omega_{p}}}\left[\frac{3\pi(\omega_{k}+\omega_{p})b}{2\omega_{p0}}\right]^{i\omega_{+}^{r}D}
\\
&\quad\times\text{exp}\left(\frac{i\pi}{4}+\frac{3i\pi(\omega_{k}+\omega_{p})b}{4\omega_{p0}}-\frac{\pi\omega_{+}^{r}D}{2}\right)\biggl[\Gamma\left(\frac{1}{2}-i\omega_{+}^{r}D\right)-\Gamma\left(\frac{1}{2}-i\omega_{+}^{r}D,\frac{3i\pi(\omega_{k}+\omega_{p})b}{2\omega_{p0}}\right) \biggr], 
\end{aligned} \label{beta trajectory 2 exact}
\end{equation}
\end{widetext}
where the complete gamma function dominates over the upper incomplete one when $(\omega_{k}+\omega_{p})b\gg \omega_{p0}$.

Since the second case with $b=1$ is favored in the proposed experiment \cite{Chen2020}, the corresponding analog Hawking spectrum appears when $\omega_{p}\sim p_3$, $\omega_{p}\gg\omega_{k}$, $\omega_{p}\gg \omega_{p0}$. Under these limits, one obtains
\begin{equation}
\begin{aligned}
&|\beta_{\textbf{k}\textbf{p}}^{(1),ref}(p_3>0)|^2
\\
&\approx\frac{A^2}{16\pi^4}\;\text{sinc}^2\left[\frac{(k_1+p_1)L}{2}\right]\text{sinc}^2\left[\frac{(k_2+p_2)L}{2}\right]
\\
&\quad\times\frac{\alpha^2D}{4\pi\omega_{k}(p_3)^2}\left[\frac{1}{e^{\omega_{k}/T_{\text{eff}(\theta_k)}}+1}\right],
\end{aligned}
\end{equation}
which is of the same form as Eq.~\eqref{beta squared trajectory 1} but with $1/(2D)$ playing the role of $\kappa$. That is, in the proposed experiment, the analog Hawking temperature is related to the inverse of the characteristic length $D$ of the plasma density gradient. In addition, the corresponding particle spectrum will also have the same expression as Eq.~\eqref{Particle spectrum trajectory 1} but with $\kappa$ replaced by $1/(2D)$ and $P_3$ determined by $p_3\gg \alpha$, $p_3\gg \omega_{k}$, and $p_3\gg \omega_{p0}$.

Since the particle spectrum in this example is identical to the previous example, the discussions in the previous subsection directly carries over. In addition, by using the parameter values: $b=1$, $D=2.5\text{ eV}^{-1}$ and $\omega_{p0}=0.006\text{ eV}$, while other parameters having the same values as those of the previous subsection's, the quantitative results of the previous subsection also apply here.

In the proposed AnaBHEL experiment \cite{Chen2020}, the mirror is to be created through laser-plasma interactions and the mirror would have a finite instead of zero thickness with SRLD density profile. Invoking a state-of-the-art petawatt-class laser facility that provides 1 laser shot per minute and 8 hours of operation time per day, a 20-day experimental data acquisition would give the total yield of Hawking events
\begin{equation}
    N_{\text{total}}=(1\times 60\times 8\times 20)\times 1\times 0.29\sim 3000.
\end{equation}

\section{Partner particle}\label{Partner particle}

In the previous sections, we only discussed about the creation of analog Hawking particles, which is sufficient enough for experiments aiming to testify the existence and distribution of such particles. Nonetheless, to further tackle the information loss paradox, it is inevitable to study the entanglement between the analog Hawking particle and its partner particle. Thus, it is also necessary to study how the partner particles distribute.

For typical quantum field theory in curved spacetimes with well-defined asymptotic free field regimes, and the moving mirror model in flat spacetime, the in-vacuum $\left|0;\text{in}\right>$ is a squeezed vacuum state, i.e.,
\begin{equation}
    \left|0;\text{in}\right>=\mathcal{Z}\text{exp}\left[\frac{1}{2}\int_{\mathbf{k,p}} \;V_{\mathbf{kp}}\hat{a}^{\text{out}\dagger}_{\mathbf{k}}\hat{a}^{\text{out}\dagger}_{\mathbf{p}}\right]\left|0;\text{out}\right>,
\end{equation}
where $\int_{\mathbf{k}}\equiv\int d^3k$, $\mathcal{Z}=\left<0;\text{out}|0;\text{in}\right>$ is the vacuum persistence amplitude, and $V_{\mathbf{kp}}=\int d^3q\;\bar{\beta}_{\mathbf{kq}}\alpha_{\mathbf{qp}}^{-1}$. In the case of fields weakly coupled to external sources, the in-vacuum becomes
\begin{equation}
\begin{aligned}
     \left|0;\text{in}\right>\approx& \mathcal{Z}\left|0;\text{out}\right>
     \\
     &+\frac{\mathcal{Z}}{2}\int_{\mathbf{k_{1},p_{1},q_{1}}}\bar{\beta}_{\mathbf{k_{1}q_{1}}}\alpha_{\mathbf{q_{1}p_{1}}}^{-1}\left|1_{\mathbf{k_{1}}},1_{\mathbf{p_{1}}};\text{out}\right>
\end{aligned}
\end{equation}
to leading non-trivial order. In addition, since $\alpha_{\mathbf{q_{1}p_{1}}}^{-1}\approx \delta(\mathbf{q_{1}}-\mathbf{p_{1}})$, which equalises the in-mode's momentum $\mathbf{q_{1}}$ to the out-mode's momentum $\mathbf{p_{1}}$, the in-vacuum further simplifies to
\begin{equation}
\begin{aligned}
     \left|0;\text{in}\right>\approx \mathcal{Z}\left[\left|0;\text{out}\right>
     +\frac{1}{2}\int_{\mathbf{k_{1},p_{1}}}\bar{\beta}_{\mathbf{k_{1}p_{1}}}\left|1_{\mathbf{k_{1}}},1_{\mathbf{p_{1}}};\text{out}\right>\right].
\end{aligned}
\end{equation}

Therefore, the in-vacuum will be found to be in either a state with zero out-particle or a state with a pair of out-particles each with momentum $\mathbf{k_{1}}$ and $\mathbf{p_{1}}$.

Since the vacuum persistence amplitude is related to the generating functional of connected Feynman diagrams $W$ through $\mathcal{Z}=e^{iW}$, the probability for the in-vacuum to transit into out-vacuum, in the case of weak coupling to external sources, is
\begin{equation}
    P(0_{\text{in}}\rightarrow 0_{\text{out}})\approx 1-2\text{Im}W,\quad 0\leq\text{Im}W\ll 1,
\end{equation}
and the probability for the in-vacuum to transit into a pair of out-particles is
\begin{equation}
    P(0_{\text{in}}\rightarrow 2_{\text{out}})=2\text{Im}W=\int d^3kd^3p\left|\beta_{\mathbf{kp}}\right|^2\ll 1,
\end{equation}
while transitions to other number of out-particles have approximately zero probabilities.

In the case of moving mirror model, the mirror plays the role of external source and the semitransparency is the weak coupling. In addition, according to the previous section, the analog Hawking spectrum corresponds to the regime in which the analog Hawking particle propagates in the positive $z$-direction, i.e., $k_3>0$, the incident in-mode propagates in the negative $z$-direction, i.e., $p_3<0$, with little transverse momentum, i.e., $\omega_{p}\sim p_{3}$, and the in-mode's frequency is much larger than the analog Hawking particle's frequency, i.e., $\omega_{p}\gg\omega_{k}$. Thus, within the pair of out-particles as final state, the one with momentum $\mathbf{k}$ is the analog Hawking particle while the other with momentum $\mathbf{p}$ is the partner particle, which has the same momentum as the incident in-mode. Moreover, the sinc functions (delta function for an infinite-size mirror) in the beta coefficient indicate that most of the partner particles have opposite transverse momenta compared to the analog Hawking particles, i.e., $\mathbf{p}_{\perp}=-\mathbf{k}_{\perp}$. Despite the analog Hawking and its partner particle having transverse momenta of the same magnitude, the large longitudinal momentum of the partner particle makes its propagation much more longitudinal compared to the analog Hawking particle. That is, while the analog Hawking particles can have a wide angular distribution, the partner particles are mostly emitted in a narrow solid angle $(\theta_{p}\leq \theta_{*} \ll 1)$ in the opposite longitudinal direction. Lastly, since the out-particles always appear in pair, the number of partner particles will be the same as that of the analog Hawking particles.

\section{Conclusion}\label{Conclusion}

In this paper, we studied the particle production by a relativistic semitransparent mirror of arbitrary potential $V$ in (1+3)D flat spacetime. In particular, we apply the derived spectrum formula to the case of a square mirror of finite transverse area $A=L\times L$ with or without longitudinal thickness.

The particle spectrum formula is applied to two specific trajectories. One is the analog gravitational collapse trajectory \eqref{trajectory 1} that models the formation of a black hole from a collapsing null shell in (1+3)D curved spacetime, and the other is the Chen-Mourou trajectory \eqref{trajectory 2}, which is proposed to be realizable in future experiment, that has a similar late time behavior as Eq.~\eqref{trajectory 1}.

The frequency and angular spectra in the parameter regime where the Hawking radiation analog can be made are studied in this paper and they are plotted in Figs.~\ref{freqangular}$-$\ref{freq}. The finite-size effects on the spectra are also clearly distinguished in these figures. Specifically, finite transverse size leads to diffraction, while nontrivial longitudinal thickness leads to enhancement of particle production. While our discussion in the main text is devoted to a semitransparent mirror, the form factors should be valid for mirrors of arbitrary reflectivity and motion since the form factors are independent of $\alpha$ and $Z(t)$.

Most importantly, we considered an experimental design, which is planned for future experiments, for which we obtained an encouraging estimation of the produced analog Hawking particles of roughly $3000$ events for a 20-day detection.

\begin{acknowledgments}
The authors appreciate helpful discussions with Yung-Kun Liu of National Taiwan University. This work is supported by ROC (Taiwan) Ministry of Science and Technology (MOST), National Center for Theoretical Sciences (NCTS), and Leung Center for Cosmology and Particle Astrophysics (LeCosPA) of National Taiwan University. P.C. is in addition supported by U.S. Department of Energy under Contract No. DE-AC03-76SF00515. 
\end{acknowledgments}

\appendix

\section{Exact solution in (1+1)D}\label{A1}

According to the general approach in Ref.~\cite{Nicolaevici2001} for a moving mirror in (1+1)D flat spacetime, we find the exact reflection coefficient on the mirror's, which follows the trajectory \eqref{trajectory 1}, right-hand side is
\begin{widetext}
\begin{align}
R^{\text{out}}_{R}(0<v<\frac{1}{\kappa})&=e^{\frac{\alpha\tau}{2}}\left(\frac{\alpha\tau}{2}\right)^{-\frac{2i\omega}{\kappa}}\left[\Gamma\left(1+\frac{2i\omega}{\kappa},\frac{\alpha\tau}{2}\right)-\Gamma\left(1+\frac{2i\omega}{\kappa}\right) \right],
\\
R^{\text{out}}_{R}(-\infty<v<0)&=\left[1-e^{(\frac{\alpha}{2}+i\omega )v}\right]\frac{\alpha}{\alpha+2i\omega}+e^{\frac{\alpha\tau}{2}}\left(\frac{\alpha\tau_0}{2}\right)^{-\frac{2i\omega}{\kappa}}\left[\Gamma\left(2+\frac{2i\omega}{\kappa},\frac{\alpha\tau_0}{2}\right)-\Gamma\left(2+\frac{2i\omega}{\kappa}\right) \right],
\end{align}
\end{widetext}
where $\tau=-(2/\kappa)\sqrt{1-\kappa v}$ for $0<v<\kappa^{-1}$ and $\tau=\tau_0+v$ with $\tau_0=-2/\kappa$ for $-\infty<v<0$.

In the case of $\alpha\ll\kappa$, the resulting reflection coefficients and beta coefficient are
\begin{widetext}
\begin{align}
R^{\text{out}}_{R}(0<v<\frac{1}{\kappa})&=\frac{\alpha}{\kappa}\frac{\kappa}{\kappa+2i\omega}\sqrt{1-\kappa v}+O\left(\frac{\alpha^2}{\kappa^2} \right),
\\
R^{\text{out}}_{R}(-\infty<v<0)&=\left[1-e^{(\frac{\alpha}{2}+i\omega )v}\right]\frac{\alpha}{\alpha+2i\omega}+\frac{\alpha}{\kappa}e^{(\frac{\alpha}{2}+i\omega )v}\frac{\kappa}{\kappa+2i\omega}+O\left(\frac{\alpha^2}{\kappa^2}\right),
\\
\beta_{\omega\omega'}^{ref}&=-\frac{\alpha}{\kappa}\frac{ie^{-\frac{i\omega'}{\kappa}}e^{\frac{i\pi}{4}}e^{-\frac{\pi\omega}{2\kappa}}}{4\pi\sqrt{\omega\omega'}}\left(\frac{\kappa}{\omega'}\right)^{\frac{1}{2}+\frac{i\omega}{\kappa}} \left[\Gamma\left(\frac{1}{2}+\frac{i\omega}{\kappa} \right)-\Gamma\left(\frac{1}{2}+\frac{i\omega}{\kappa} ,-\frac{i\omega'}{\kappa}\right) \right]+\frac{\omega'}{2\pi\sqrt{\omega\omega'}}\frac{\alpha}{\kappa}\frac{\kappa}{\kappa+2i\omega}\frac{i}{\omega'}\nonumber
\\
&\quad -\frac{\omega'}{2\pi\sqrt{\omega\omega'}}\left[\frac{\alpha}{\alpha+2i\omega}\frac{i}{\omega+\omega'}-\frac{\alpha}{\alpha+2i\omega}\frac{2}{\alpha-2i\omega'}+\frac{\alpha}{\kappa}\frac{\kappa}{\kappa+2i\omega}\frac{2}{\alpha-2i\omega'} \right]+O\left(\frac{\alpha^2}{\kappa^2}\right),
\label{A5}
\end{align}
\end{widetext}
which are valid for $\omega\in[0,\infty)$ and $\omega'\in[0,\infty)$. By further taking the limits: $\alpha\ll\omega$ and $\alpha\ll\omega'$, we have
\begin{widetext}
\begin{align}
\begin{aligned}
\beta_{\omega\omega'}^{ref}&=-\frac{\alpha}{\kappa}\frac{ie^{-\frac{i\omega'}{\kappa}}e^{\frac{i\pi}{4}}e^{-\frac{\pi\omega}{2\kappa}}}{4\pi\sqrt{\omega\omega'}}\left(\frac{\kappa}{\omega'}\right)^{\frac{1}{2}+\frac{i\omega}{\kappa}} \left[\Gamma\left(\frac{1}{2}+\frac{i\omega}{\kappa} \right)-\Gamma\left(\frac{1}{2}+\frac{i\omega}{\kappa} ,-\frac{i\omega'}{\kappa}\right) \right]
\\
&\quad+\frac{\omega'}{2\pi\sqrt{\omega\omega'}}\frac{\alpha}{2\omega'(\omega+\omega')}+O\left(\frac{\alpha^2}{\kappa^2}\right)+O\left(\frac{\alpha^2}{\omega^2}\right)+O\left(\frac{\alpha}{\omega'}\right),
\end{aligned}
\end{align}
\end{widetext}
which reproduces our previous result for a semitransparent mirror in the (1+1)D limit \cite{Lin2020}.

The analog Hawking particles can be extracted by further taking the approximations: $\omega'\gg\kappa$ and $\omega'\gg\omega$, which gives
\begin{equation}
\beta_{\omega\omega'}^{ref}\approx-\frac{\alpha}{\kappa}\frac{ie^{-\frac{i\omega'}{\kappa}}e^{\frac{i\pi}{4}}e^{-\frac{\pi\omega}{2\kappa}}}{4\pi\sqrt{\omega\omega'}}\left(\frac{\kappa}{\omega'}\right)^{\frac{1}{2}+\frac{i\omega}{\kappa}} \Gamma\left(\frac{1}{2}+\frac{i\omega}{\kappa} \right).
\label{A7}
\end{equation}

On the other hand, taking the approximations: $\omega\ll\alpha\ll\kappa\ll\omega'$ starting from Eq.~\eqref{A5} also leads to
 \begin{equation}
 \begin{aligned}
\beta_{\omega\omega'}^{ref}&\approx-\frac{\alpha}{\kappa}\frac{ie^{-\frac{i\omega'}{\kappa}}e^{\frac{i\pi}{4}}e^{-\frac{\pi\omega}{2\kappa}}}{4\pi\sqrt{\omega\omega'}}\left(\frac{\kappa}{\omega'}\right)^{\frac{1}{2}+\frac{i\omega}{\kappa}} \Gamma\left(\frac{1}{2}+\frac{i\omega}{\kappa} \right).
\end{aligned}
\label{A8}
\end{equation}

Therefore, whether $\omega\ll\alpha$ or $\alpha\ll\omega$, which is the case in the perturbative approach using the Green functions, there are identical beta coefficients, i.e., Eqs.~\eqref{A7} and \eqref{A8}, for the analog Hawking particles.

\section{Transition amplitude}\label{Transition amplitude}

In this section, we provide another derivation of the particle spectrum. Suppose the initial state is $\left|0;\text{in}\right>$ and the final state is $\left|\mathbf{p},\mathbf{k};\text{out}\right>=\hat{a}^{\text{out}\dagger}_{\mathbf{p}}\hat{a}^{\text{out}\dagger}_{\mathbf{k}}\left|0;\text{out}\right>$, where the in- and out-vacua are related by the scattering matrix $\mathbb{S}$ by $\left<0;\text{out}\right|=\left<0;\text{in}\right|\mathbb{S}$. Then according to the Bogoliubov transformation of creation and annihilation operators, we find
\begin{align}
    \frac{\left<\mathbf{p},\mathbf{k};\text{out}|0;\text{in}\right>}{\left<0;\text{out}|0;\text{in}\right>}=\beta_{\mathbf{kp}}^{(1)}+\mathcal{O}(\alpha^2).
\end{align}
Therefore, the number of out-particles in the mode $\mathbf{k}$, i.e., the average occupancy number, can also be expressed as
\begin{equation}
\frac{dN}{d^3k}=\int d^3p\left| \frac{\left<\mathbf{p},\mathbf{k};\text{out}|0;\text{in}\right>}{\left<0;\text{out}|0;\text{in}\right>} \right|^2.
\end{equation}

The transition amplitude can be computed by the Lehmann-Symanzik-Zimmermann (LSZ) reduction formula:
\begin{widetext}
\begin{align}
   \frac{\left<\mathbf{p},\mathbf{k};\text{out}|0;\text{in}\right>}{\left<0;\text{out}|0;\text{in}\right>}&=\frac{i}{[2\pi]^{3/2}[2\omega_{k}]^{1/2}}\frac{i}{[2\pi]^{3/2}[2\omega_{p}]^{1/2}}\int_{\mathbb{R}} d^4x\; e^{-ip\cdot x}\left(-\partial^{\mu}\partial_{\mu}+m^2 \right)\int_{\mathbb{R}} d^4y\; e^{-ik\cdot y}\left(-\partial^{\nu}\partial_{\nu}+m^2 \right)\tau(x,y),
\end{align}
\end{widetext}
where $m$ is the field mass and
\begin{align}
    \tau(x,y)=\frac{\int D\phi\; \phi(x)\phi(y)e^{iS[\phi]}}{\int D\phi\; e^{iS[\phi]}},
\end{align}
where $S[\phi]$ is the full action of the theory.

Insertion of the action \eqref{action} into the above formulae gives
\begin{align}
     &\frac{\left<\mathbf{p},\mathbf{k};\text{out}|0;\text{in}\right>}{\left<0;\text{out}|0;\text{in}\right>}\nonumber
     \\
     &\approx\frac{-i\alpha}{8\pi^3[2\omega_{k}]^{1/2}[2\omega_{p}]^{1/2}}
     \int_{\mathbb{R}} d^4x' V(x')e^{-i(k+p)\cdot x'},
\end{align}
which reproduces Eq.~\eqref{beta} in the massless limit.

\section{Diffraction of mode function}\label{Diffraction of mode function}

Before advancing to the applications of the previously derived beta coefficient, let us pause for a second to examine the diffraction phenomenon encoded in the mode function in the case of a finite-size plane mirror. 

For simplicity, let us consider the case of a left-moving incident plane wave $u_{\text{inc}}=\text{exp}[-i\omega(t+z)]$ scattering with the semitransparent mirror via the potential \eqref{Finite plane mirror}. According to Eq.~\eqref{Field first order in}, the scattered wave is then
\begin{align}
&u_{\text{scattered}}(x)\label{Diffracted wave 1}
\\
&=-\alpha\int_{\mathbb{R}}d^4x'G_{R}(x,x')V(x')u_{\text{inc}}(x')\nonumber
\\
&=-\alpha\int_{\mathbb{R}}dt'\gamma^{-1}(t')e^{-i\omega t'-i\omega Z(t')}\int_{\mathbb{R}}\frac{d\omega'}{2\pi}\nonumber
\\
&\quad\times\frac{e^{-i\omega'(t-t')}}{4\pi}\int_{-L/2}^{L/2}d^2x_\perp'\left.\frac{e^{i\omega'|\textbf{x}-\textbf{x}'|}}{|\textbf{x}-\textbf{x}'|}\right|_{z'=Z(t')}.\nonumber
\end{align}

In the case of a mirror at rest at $Z(t')=0$, the scattered wave in the far-field regime $(|z|\gg x_{\perp},\;|z|\gg L)$ then becomes
\begin{equation}
\begin{aligned}
&u_{\text{scattered}}(x)
\\
&\approx-\frac{\alpha A e^{-i\omega(t-|\mathbf{x}|)}}{4\pi |\mathbf{x}|}\;\text{sinc}\left(\frac{\omega L x}{2|\mathbf{x}|}\right)\text{sinc}\left(\frac{\omega L y}{2|\mathbf{x}|}\right),
\end{aligned}\label{Diffracted wave 2}
\end{equation}
where we have used
\begin{equation}
|\textbf{x}-\textbf{x}'|\approx|\mathbf{x}|-\frac{xx'+yy'+zz'}{|\mathbf{x}|}
\end{equation}
before evaluating the integrations over $x'$ and $y'$. The diffraction phenomenon is encoded in the Sinc functions. More general situations such as an incident wave with transverse momentum scattering with a relativistically moving mirror can also be computed via Eq.~\eqref{Field first order in}. However, it is already sufficient to convince oneself that the diffraction is indeed encoded in the formula by the simple example we demonstrated above.

From Eq.~\eqref{Diffracted wave 2}, it is also straightforward to obtain the differential cross section $(d\sigma/d\Omega)$ by excluding the factor of spherical wave and then taking the modulus square. The result is
\begin{equation}
\begin{aligned}
\frac{d\sigma}{d\Omega}&\approx\frac{\alpha^2A^2}{16\pi^2}\;\text{sinc}^2\left(\frac{\omega L x}{2|\mathbf{x}|}\right)\text{sinc}^2\left(\frac{\omega L y}{2|\mathbf{x}|}\right).
\end{aligned}
\end{equation}

As a parallel comparison to classical scalar diffraction theory, let us consider the case of a left-moving incident plane wave $u_{\text{inc}}=\text{exp}[-i\omega(t+z)]$ scattering with a perfectly reflecting plane mirror at rest at $Z(t')=0$. This can be achieved by considering a wave $u(x)$ obeying the Klein-Gordon (KG) equation:
\begin{equation}
\partial^{\mu}\partial_{\mu}u(x)=0,\quad\mu=0,1,2,3,\label{KG equation}
\end{equation}
and subject to the mixed boundary condition:
\begin{equation}
\begin{cases}
u(x)&=0,\quad\text{for }z=0,\{x,y\}\in S
\\
u(x)&=u_{\text{inc}}(x),\quad\text{for }z=0,\{x,y\}\in \bar{S},
\end{cases}
\end{equation}
where $S=[-L/2,+L/2]$ and $\bar{S}$ is the complement of $S$. The solution to this problem is
\begin{equation}
\begin{aligned}
&u(x)=u_{\text{inc}}(x)
\\
&+2\int_{\mathbb{R}} dt'\int_{-L/2}^{L/2}d^2x_{\perp}'u_{\text{inc}}(x')\partial_{3}G_{R}(x,x'),
\end{aligned}\label{Diffracted perfect 1}
\end{equation}
where $z'$ is evaluated at $z'=0$. To check that it is indeed a solution to the KG equation and satisfies the mixed boundary condition, it is more convenient to use Eq.~\eqref{R/A green function momentum space} for the retarded Green function in Eq.~\eqref{Diffracted perfect 1}. The result is
\begin{equation}
\begin{aligned}
&u(x)=u_{\text{inc}}(x)
\\
&-\frac{e^{-i\omega t}}{4\pi^2}\int_{-L/2}^{L/2}d^2x_{\perp}'\int_{\mathbb{R}}d^2k_{\perp}e^{i\mathbf{k}_{\perp}\cdot(\mathbf{x}_{\perp}-\mathbf{x}_{\perp}')}e^{i\sqrt{\omega^2-k_\perp^2}z}.
\end{aligned}\label{Diffracted perfect 2}
\end{equation}

It can now be easily checked that Eq.~\eqref{Diffracted perfect 2} satisfies the KG equation by inserting Eq.~\eqref{Diffracted perfect 2} into Eq.~\eqref{KG equation}. Furthermore, when $z=0$, the integration over $\mathbf{k}_\perp$ can be performed to give $(2\pi)^2\delta(\mathbf{x}_\perp-\mathbf{x}_\perp')$. On one hand, if $\{x,y\}\in S$, the scattered wave becomes $\text{exp}(-i\omega t)$, which then cancels with the incident wave leading to $u(x)=0$. On the other hand, if $\{x,y\}\in \bar{S}$, the scattered wave vanishes since $\{x',y'\}\in S$ and one obtains $u(x)=u_{\text{inc}}(x)$.

To appreciate the diffraction phenomenon that appears in the standard scalar diffraction theory, it is, however, more convenient to use the expression of the retarded Green function employed in Eq.~\eqref{Diffracted wave 1}. In this manner, the scattered wave becomes
\begin{equation}
\begin{aligned}
&u_{\text{scattered}}(x)
\\
&=\frac{e^{-i\omega t}}{2\pi}\int_{-L/2}^{L/2}d^2x_{\perp}'\frac{z e^{i\omega|\textbf{x}-\textbf{x}'|}}{|\textbf{x}-\textbf{x}'|^2}\left[i\omega-\frac{1}{|\textbf{x}-\textbf{x}'|}\right]
\end{aligned}
\end{equation}
with $z'$ evaluated at $z'=0$. In the far-field regime $(z\gg x_{\perp},\;z\gg L)$ and for incident wave with wavelength small compared to the mirror-observation point distance, i.e., $\lambda=2\pi/\omega\ll |\textbf{x}-\textbf{x}'|$, the scattered wave simplifies to
\begin{equation}
\begin{aligned}
&u_{\text{scattered}}(x)
\\
&\approx\frac{izAe^{-i\omega(t-|\mathbf{x}|)}}{\lambda|\mathbf{x}|^2}\;\text{sinc}\left(\frac{\omega L x}{2|\mathbf{x}|}\right)\text{sinc}\left(\frac{\omega L y}{2|\mathbf{x}|}\right).
\end{aligned}\label{Diffracted perfect 3}
\end{equation}

For both semitransparent and perfectly reflecting static mirrors, diffraction manifests itself in the Sinc functions. Nevertheless, interestingly, aside from the common Sinc functions in Eqs.~\eqref{Diffracted wave 2} and \eqref{Diffracted perfect 3}, their prefactors are also similar. Indeed, since $\alpha\sim 1/\lambda_{\text{IR}}$, where $\lambda_{\text{IR}}$ is the wavelength of the infrared cutoff due to the mirror's semitransparency, Eq.~\eqref{Diffracted wave 2} can be cast into the same form as Eq.~\eqref{Diffracted perfect 3}.

\section{Time-dependent spectrum}\label{Time-dependent spectrum}

In the previous sections, to obtain the analog Hawking spectrum, we only focused on the frequency/momentum regimes: $\omega_{p}\sim p_3$, $\omega_{p}\gg\omega_{k}$, and $\omega_{p}\gg\kappa$ or $\omega_{p0}$. In this section, we shall demonstrate that field modes in these regimes are the dominant contributions to what an observer would observe at late times.

The Bogoliubov coefficients, and therefore the particle spectra, in the previous sections are non-local quantities, i.e., they are not functions of spacetime coordinates and they depend on the mirror's entire history. Therefore, it is not clear what an observer in the out-region would see as time evolves. The study of time dependency can be achieved by treating the field modes as wave packets \cite{Hawking1975}.

Since we are interested in what an out-observer would see as time evolves, we consider the wave packet:
\begin{equation}
    v_{jn}(x)\equiv\frac{1}{\sqrt{\epsilon}}\int_{j\epsilon}^{(j+1)\epsilon}d\omega_{k}\;e^{-\frac{2\pi i\omega_{k}n}{\epsilon}}v_{\mathbf{k}}(x),
\end{equation}
where $v_{\mathbf{k}}(x)$ is the out-mode whose explicit form is irrelevant but $\lim_{t\rightarrow\infty}v_{\mathbf{k}}(x)=v_{\mathbf{k}}^{\text{out}}(x)$, $\epsilon>0$ is the frequency bin width, e.g., allowed by a particle detector, and $j\geq 0$ and $n$ plays the role of time. The time-dependent Bogoliubov beta coefficient is then
\begin{equation}
    \beta_{jn\mathbf{p}}=\frac{1}{\sqrt{\epsilon}}\int_{j\epsilon}^{(j+1)\epsilon}d\omega_{k}\;e^{-\frac{2\pi i\omega_{k}n}{\epsilon}}\beta_{\mathbf{kp}}.
\end{equation}

In the case of the frequency bin width being narrow, i.e., $\epsilon\rightarrow 0$, and using Eq.~\eqref{beta trajectory 1 exact}, we obtain
\begin{widetext}
    \begin{align}
    \beta_{jn\mathbf{p}}^{(1),ref}\approx&\frac{A}{4\pi^2}\;\text{sinc}\left[\frac{(k_1+p_1)L}{2}\right]\text{sinc}\left[\frac{(k_2+p_2)L}{2}\right]\left[\frac{-\alpha\sqrt{\epsilon}}{4\pi \sqrt{\omega_{k}}\sqrt{\omega_{p}}} \right]\biggl\{\frac{e^{i\omega_{k}\eta_{1}}}{\omega_{k}+\omega_{p}}\; \text{sinc}\left(\frac{\epsilon \eta_{1}}{2}\right)+\frac{1}{\kappa}\left(\frac{2\kappa}{\omega_{-}^{r}}\right)^{\frac{1}{2}-\frac{i(\omega_{p}-p_3)}{2\kappa}}
    \nonumber
    \\
    &\times\text{exp}\left(\frac{i\pi}{4}+\frac{i\omega_{-}^r}{2\kappa}-\frac{\pi \omega_{+}^r}{4\kappa}\right)\biggl[\Gamma\left(\frac{1}{2}-\frac{i\omega_{+}^r}{2\kappa} \right)-\Gamma\left(\frac{1}{2}-\frac{i\omega_{+}^r}{2\kappa},\frac{i\omega_{-}^r}{2\kappa}\right) \biggr]e^{i\omega_{k}\eta_{2}}\text{sinc}\left(\frac{\epsilon \eta_{2}}{2}\right)
    \biggr\},
    \label{beta trajectory 1 exact WP}
    \end{align}
\end{widetext}
where $k_{1}=\omega_{k}\sin\theta_{k}\cos\phi_{k}$, $k_{2}=\omega_{k}\sin\theta_{k}\sin\phi_{k}$, $\eta_{1}=-2\pi n/\epsilon$, $\eta_{2}=(2\kappa)^{-1}(1+\cos\theta_{k})\ln[\omega_{-}^{r}/(2\kappa)]-2\pi n/\epsilon$, and all out-mode frequencies are approximated by the center value of the bin width, i.e., $\omega_{k}\approx(j+1/2)\epsilon$.

Similarly, for $\epsilon\rightarrow 0$ and using Eq.~\eqref{beta trajectory 2 exact}, one obtains
\begin{widetext}
    \begin{align}
    \beta_{jn\mathbf{p}}^{(1),ref}\approx&\frac{A}{4\pi^2}\;\text{sinc}\left[\frac{(k_1+p_1)L}{2}\right]\text{sinc}\left[\frac{(k_2+p_2)L}{2}\right]\left[\frac{-\alpha\sqrt{\epsilon}}{4\pi \sqrt{\omega_{k}}\sqrt{\omega_{p}}} \right]\frac{\sqrt{2D}}{\sqrt{\omega_{k}+\omega_{p}}}\;\text{exp}\left(\frac{i\pi}{4}+\frac{3i\pi(\omega_{k}+\omega_{p})b}{4\omega_{p0}}-\frac{\pi\omega_{+}^{r}D}{2}\right)
    \nonumber
    \\
    &\times\biggl[\Gamma\left(\frac{1}{2}-i\omega_{+}^{r}D\right)-\Gamma\left(\frac{1}{2}-i\omega_{+}^{r}D,\frac{3i\pi(\omega_{k}+\omega_{p})b}{2\omega_{p0}}\right) \biggr]\left[\frac{3\pi(\omega_{k}+\omega_{p})b}{2\omega_{p0}}\right]^{i(\omega_{p}-p_3)D}e^{i\omega_{k}\eta_{3}}\text{sinc}\left(\frac{\epsilon \eta_{3}}{2}\right),
    \label{beta trajectory 2 exact WP}
    \end{align}
\end{widetext}
where $\eta_{3}=(1+\cos\theta_{k})D\ln[3(\omega_{k}+\omega_{p})b/(2\omega_{p0})]-2\pi n/\epsilon$, and, as before, $\omega_{k}\approx (j+1/2)\epsilon$.

At late times, i.e., $n\rightarrow\infty$, one sees that $\eta_{1}\rightarrow -\infty$, $\eta_{2}\rightarrow-\infty$ unless $\omega_{-}^{r}\gg\kappa$ such that $\eta_{2}=0$, and $\eta_{3}\rightarrow-\infty$ unless $(\omega_{k}+\omega_{p})b\gg\omega_{p0}$ such that $\eta_{3}=0$. Therefore, at late times, only the complete gamma function terms in the time-dependent beta coefficients give non-vanishing contributions. In addition, since the mirror is receding at near-the-speed-of-light at late times, we know that only in-modes with $\omega_{p}\sim p_{3}$ can catch up the mirror and interact with it. These lead to the main contribution to the spectrum coming from $\omega_{k}\sim \kappa/(2\pi)$ or $(4\pi D)^{-1}$. The above altogether implies that, at late times, the time-dependent beta coefficient \eqref{beta trajectory 1 exact WP} would be dominated by $\omega_{p}\sim p_{3}\gg\kappa\gtrsim\omega_{k}$, and Eq.~\eqref{beta trajectory 2 exact WP} would be dominated by $\omega_{p}\sim p_{3}$, $\omega_{k}\lesssim D^{-1}$, and $(\omega_{k}+\omega_{p})b\gg\omega_{p0}$. In the previous sections, we used $D=2.5$ eV$^{-1}$, $\omega_{p0}=0.006$ eV and $b=1$, so $\omega_{p}\sim p_{3}\gg \omega_{p0}\gtrsim\omega_{k}$.

By considering the out-modes as wave packets, we explicitly demonstrated that the frequency/momentum approximations made in the previous sections to obtain the analog Hawking spectrum correspond to what an out-observer would detect at late times.

\nocite{*}

\bibliography{Particle_production_by_a_relativistic_semitransparent_mirror_of_finite_size_and_thickness}

\end{document}